\documentstyle[12pt,epsf]{article}
\newcommand{\be}{\begin{equation}}
\newcommand{\ee}{\end{equation}}
\newcommand{\abz}{\hspace*{.5in}}

\begin{document}

\title{
THERMODYNAMIC STABILITY OF A MULTI-BUBBLE
 COSMOLOGICAL MODEL\thanks{talk at Seventh Marcel
Grossman Meeting, Stanford University, July 1994}}
\author{GERALD HORWITZ\thanks{Supported by BSF research
grant No. 89-00244} and OLEG FONAREV
\\ {\em Racah Institute of Physics, The Hebrew University}
\\ {\em Jerusalem 91904, Israel}}
\date{}

\maketitle
\begin{abstract}
Multibubble solutions for a cosmological model which
lead to thermal inflationary states due to a semi-classical
tunneling of gravity are calculated.
\end{abstract}
\section{One domain solutions}
\abz In earlier work done by several different approaches \cite{HW,RB} ,  an
inflationary
solution was found as semiclassical tunneling using a simple model of
quantum matter and semiclassical gravity.   The model comprised massless
noninteracting scalar bosons conformally coupled to gravity in a RW universe
with $k=+1$ and a positive phenomenological  cosmological constant $\Lambda$.
 For such model there is a classically forbidden region connecting two
classically allowed regions. The inner one of these two is inappropriate
for our model since  the inner region necessarily involves quantum gravity.
 The model was evaluated by two different approaches, first by an explicit
statistical mechanics approach and then by a wave function of the universe
 approach.   In both of these approaches the initial method was to calculate
  the solutions  in the outer
classically allowed region.  The result of tunneling is to produce an
inflationary thermal solution in the outer region.  This tunneling yields
both a definition of physical time and entropy appropriate to an
 inflationary solution. \\
\abz We use a RW metric written in conformal time coordinates:
\be
ds^2=a^2(u)[du^2-d \chi^2 -\sin^2 \chi d\Omega^2]=a^2(u) \gamma_{\mu \nu}
dx^{\mu} dx^{\nu}
\ee
 \abz The action obtained after going over to a Euclideanized metric as
appropriate
for a tunneling region is
\be
I=\int{ dx^4 [\frac{3}{8 \pi} (\dot{a}^2 + a^2 - \Lambda/3 a^4) +\frac{1}{2}
{(\gamma^{\mu
\nu} \phi_{,\mu} \phi_{,\nu} -\phi^2)}}]=I_G + I_M
\ee
after carrying out a conformal transformation $g=a^2\gamma$
and
$\phi=a \varphi$,  whence $R/6 =1$.
One then evaluated the norm of the wave function of the universe and
showed that
\be
{P}(a) = \int d\phi |\Psi(\phi,a) |^2 = \int \frac{d\beta'}{2 \pi i}
\int_{WKB} D[a] \int D[\phi] e^{-{I}(a,\phi,\beta';a,\phi,0)}
\ee
where $\phi$ and $a$ are periodic functions of $\beta'$.  The $\beta'$
integral
fixes the total energy to be zero and its saddle point value is the
reciprocal
temperature when we have exchanged the order of integration as in the
second
part of equation (3).   The saddle point evaluation of  $a$  involves
replacing
$a$ in the integrand  by its saddle point value $\bar{a}$.
The $\phi$ functional integral gives
\be
\int D[\phi] e^{-{I}(\phi,\beta)} = e^{-\beta {F}(\beta)}
\ee
where ${F}(\beta)$ is the Helmholtz free energy.
Then ${P}(\bar{a})=e^{S}=e^{S_{G}+S_{M}}$ ,
where $S$ is the total entropy and $S_{G}$ and $S_{M}$ are respectively  the
entropies of the gravity and of the  matter.
The common saddle point solutions of the $a$ and $\beta'$ integrations
leads to the equation
\be
\beta = 2 \int_{x_{-}}^{x_{+}} dx \frac{1}{\sqrt{-e+x^2-x^4}}
\ee
where $x=(\frac{\Lambda}{3})^{1/2} a$ and  $e$  is the thermal average energy
multiplied by $\frac{\Lambda}{3}$;
$e=\frac{8}{3} \pi \Lambda k_{B} \beta^{-4}= \alpha \beta^{-4}$. \\

\begin{figure}[t]
\hspace{-6cm}
   \rightline{\epsfxsize = 7cm \epsffile{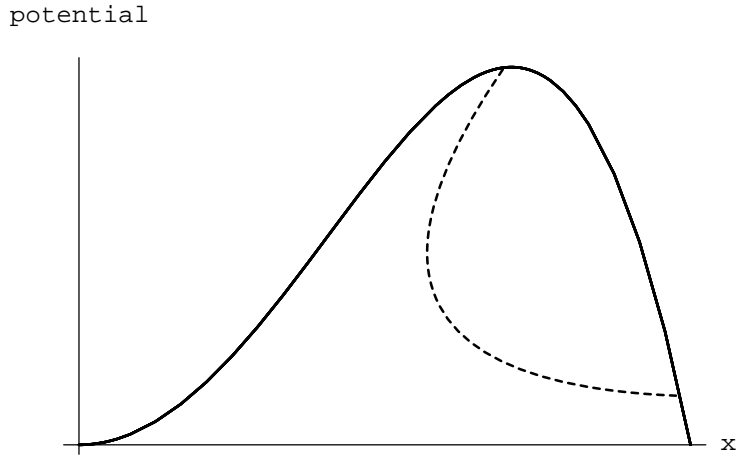}}
\vspace{-35mm}
\caption{Solutions of equation (5)}
\end{figure}

Seeking the self consistent solution in the tunneling region one finds
two
solutions in a certain range of the values of $x$ which can most clearly
be
seen by the following graph, where the dashed curve is the locus of
solutions.
Each point of the dashed line is the end point of a straight line from
$x_{-}$
at that energy.  Thus we see of all there are no solutions  for $x$ less
than
some minimum value $x_{0}$ and that there are two branches of the
solutions for
$x_{0} < x < x_{M}=1/\sqrt{2}$.
Both branches correspond to an increasing entropy as a function of $x$ and
both
are maximum entropy states of nearly equal weight near the maximum value
of the
upper solution.  The upper solution joins the real time domain where it
corresponds to a static Einstein solution.   The lower one emerges at
higher
entropy and corresponds  to a thermal inflationary solution.  Note that in
this
case there is no need  for time to establish  the thermal solutions, since
it begins in thermal equilibrium when it emerges in the real time domain. \\
\abz In previous work  we discarded the upper solution since it is of lower
entropy
and is also unstable. Subsequently we began to consider whether it is
justified
to ignore alternative saddle point solutions which cannot immediately be
discarded as being of much less weight.  The present paper is an effort to
consider another alternative.  \\

\section{Multidomain solutions}
\abz In the standard functional integral approach, if there is more than one
 saddle point the wave function in lowest order  is  a superposition of the
several contributions.  This method is not viable in our case since
we are considering thermally mixed states. We apply instead another approach in
which the superposition consists of the different solutions cooexisting in
spatially separated regions.  We worked out in detail the case of two
domains; the multibubble solution were  treated only approximately.
Let us first consider the two bubble case.  The two states are a
Robertson-Walker (RW) inflationary universe and a static Einstein
universe (SE). The two different  solutions are matched  on a singular
surface, which is taken to be spherical. The Euclidean line elements  in
the two regions take the form:
\be
ds^{2}_{RW}=\frac{3}{\Lambda} {x^{2}}(u) [ du^{2}+d\chi^{2}_{RW}+
\sin^{2}\chi_{RW} d\Omega^{2}]
\ee
\be
ds^{2}_{SE}=\frac{3}{2\Lambda} [ {N_{SE}^{2}}(u) du^{2}+d\chi^{2}_{SE}+
\sin^{2}\chi_{SE} d\Omega^{2}]
\ee
\abz The singular hypersurface is given either by the equation
$\chi_{RW}={f_{RW}}(u)$,
or by the equation $\chi_{SE}={f_{SE}}(u)$.
The matching conditions on the hypersurface lead to the following
relations:
\be
{x}(u) \sin {f_{RW}}(u) = \frac{1}{\sqrt{2}} \sin {f_{SE}}(u)
\ee
\be
 {x^{2}}(u) (1+{\dot{f}^{2}_{RW}}(u))=\frac{1}{2}
({N_{SE}^{2}}(u)+{\dot{f}^{2}_{SE}}(u))
\ee
\abz Another  constraint follows from the junction conditions that relate  the
jump in the Einstein tensor to the surface energy \cite{IL}.  As a toy model
we assume that the surface energy is purely dynamic and it comes from the
surface term in the action. We showed elsewhere \cite{FH} that the variation of
the action with respect to the induced metric  on the hypersurface
yields,
in our case the following constraint:
\be
f_{RW} = \mbox{constant} .
\ee
\abz
The thermodynamic entropy we obtained is the contribution from the two states
\be
S= S_{RW} + S_{SE}
\ee
each contribution being the sum of  the gravitational and the matter
entropies.
\be
S_{RW}={\frac{9}{2\Lambda }}V_{RW}\left[4 \int_{x_{-}(e)}^{x_{+}(e)}
 dx\sqrt{x^2 -x^4-e}+{4 \over 3} \alpha^{1/4} e^{3/4} \right]
\ee
\be
S_{SE}(e)={9 \over {2\Lambda }}\left[ {{1 \over 4}\int_0^\beta {d\bar{u}
N_2(\bar{x})\bar V_{SE}(\bar x)+{1 \over 3}\bar V_2(\bar x_+)\alpha \beta
_2^{-3}(\bar e,\beta )}} \right]
\ee
 Here $V_{RW}=\int_0^{f_{RW}} {\sin ^2\chi \ d\chi }$ and
$V_{SE}=\int_{{f_{SE}}(u)}^\pi {\sin ^2\chi }\ d\chi$  are the volumes
 occupied by the RW and SE bubbles respectively.
 The function $\bar{x}(u)$ is the saddle point solution of the variational
 equations
 for the RW region:
\be
{\bar{x}}^2=\bar x^2-\bar x^4-\bar e
\ee
with  $\bar e$   being found from the equation
\be
\left( \frac{\alpha}{\bar e} \right)^{\frac{1}{4}}=
2\int_{x_{-}(\bar e)}^{x_{+}(\bar e)}{\frac{dx'}{\sqrt{x'^2-x'^4-\bar e}}}
\ee
The (proper) temperature in the SE region is related to $\beta$ with the lapse
function:
\be
\beta _2(\bar e,\beta )=\int_{0}^{\beta} {du  \bar N_2(\bar x)}.
\ee
The condition for the two bubbles to be in thermal equilibrium is
\be
\beta _2(\bar e,(\alpha / \bar e)^{1\over 4})=(4\alpha )^{1\over 4}
\ee
\abz Equations (8)--(10) , (15)--(17) constitute the conditions to be satisfied
in equilibrium. Having found $\bar e, f_{RW}, {\bar x}(u), {{\bar f}_{SE}}(u)$
and ${{\bar N}_{SE}}(u)$ from these equations, we can find the
value of the entropy at equilibrium and check the stability from
eqs. (11)--(13).
Analysis of eqs. (15)--(17) shows that an equilibrium can only exist
for the range of parameter $\alpha$, $\pi^{4}/4 < \alpha < \pi^{4}$.
In this range there is a critical value
$\alpha_{c}$, such that the equilibrium is stable against global fluctuations
when $\alpha < \alpha_{c}$ and is quasistable when $\alpha > \alpha_{c}$.
In either case the entropy of the two
bubble state is greater than the entropy of the single RW universe. \\
\abz We now turn to the case of multiple bubbles.  There are two
possibilities:
many RW bubbles in a SE matrix or many SE bubbles in a RW matrix. It can
be shown that conditions for equilibrium remain the same as in the case
of two bubbles.  In our simple model the number of bubbles cannot exceed
25.  Moreover, the constraints limit the size of the RW region, so that
inflation cannot last long.  We are currently exploring possible
extensions of the model to overcome these problems.

\section*{Acknowledgements}
G.H. acknowledges some useful and illuminating discussions
with J. Bekenstein.

\end{document}